\documentclass[prc,twocolumn,superscriptaddress,showpacs,amssymb,amsmath,amsfonts,aps]{revtex4}
\usepackage{graphicx}
\usepackage{dcolumn}
\usepackage{epsfig}

\begin{document}
\title{Configuration mixings and
light-front relativistic quark model 
predictions for the
electroexcitation of the
$\Delta(1232)\frac{3}{2}^+$,
$N(1440)\frac{1}{2}^+$, and
$\Delta(1600)\frac{3}{2}^+$
\\}

\newcommand*{\JLAB }{ Thomas Jefferson National Accelerator Facility, 
Newport News, Virginia 23606, USA}
\affiliation{\JLAB }
\newcommand*{\YEREVAN }{ Yerevan Physics Institute, 375036 Yerevan, 
Armenia}
\affiliation{\YEREVAN }
\author{I.G.~Aznauryan}
     \affiliation{\JLAB}
     \affiliation{\YEREVAN}
\author{V.D.~Burkert}
     \affiliation{\JLAB}
\begin{abstract}
{We investigate the impact of the configurations mixings
that follow from QCD-inspired interquark forces on our results
for the electroexcitation of the
$\Delta(1232)\frac{3}{2}^+$,
$N(1440)\frac{1}{2}^+$, and
$\Delta(1600)\frac{3}{2}^+$ obtained earlier in the 
light-front relativistic quark model. 
We have shown that the configurations mixings
increase the $3q$ contribution to the
$\gamma^* N \rightarrow \Delta(1232)\frac{3}{2}^+$
magnetic-dipole
form factor at $Q^2=0$ from $42\%$ to $63\%$
and significantly improve the agreement with experiment
for the
$\gamma^* p \rightarrow N(1440)\frac{1}{2}^+$
transverse helicity amplitude at $Q^2  > 1.5~$GeV$^2$.
For the
$\gamma^* N \rightarrow \Delta(1600)\frac{3}{2}^+$
transition, configuration mixings change strongly the results
obtained earlier for the $N$ and $\Delta(1600)\frac{3}{2}^+$
taken as pure states in the multiplets $[56,0^+]$ and 
$[56',0^+]$.}
\end{abstract}
\maketitle

\section{Introduction}
\label{intro}
Recently we have reported 
light-front relativistic quark model (LF RQM) predictions for
the $\gamma^* N \rightarrow N$ form factors and 
the $\gamma^* N \rightarrow \Delta(1232)\frac{3}{2}^+$,
$N(1440)\frac{1}{2}^+$, and
$\Delta(1600)\frac{3}{2}^+$ transitions 
\cite{Aznauryan2012,Aznauryan2015}.
The predictions were made assuming that the $N$ and
$\Delta(1232)\frac{3}{2}^+$ are pure states in the multiplet
$[56,0^+]$, and the
$N(1440)\frac{1}{2}^+$ and
$\Delta(1600)\frac{3}{2}^+$ are members of the multiplet
$[56',0^+]$. However, 
it is known that in the QCD inspired quark models, where the unknown
long-distance properties of QCD are subsumed into a confining potential,
the remaining interquark forces are assumed to be 
dominated by the one-gluon exchange \cite{Rujula,Gershtein,Isgur82,Isgur87}.
These interquark forces result in configuration
mixings. In a space corresponding to the $SU(6)$ multiplets [56], [70], and [20],
the mixings have been investigated in Refs. \cite{Gershtein,Isgur82,Isgur87},
and it was found that the 
$N$, $\Delta(1232)\frac{3}{2}^+$,
$N(1440)\frac{1}{2}^+$, and
$\Delta(1600)\frac{3}{2}^+$ are predominantly mixings of the
states $[56,0^+]$, 
$[56',0^+]$, and $[70,0^+]$:
\begin{eqnarray}
|X,3q>=a_X[56,0^+]+b_X[56',0^+]+c_X[70,0^+],
\label{eq:e1}\\
|X_r,3q>=a_{X_r}[56',0^+]+b_{X_r}[56,0^+]+c_{X_r}[70,0^+], 
\label{eq:e11}
\end{eqnarray}
where $X$ and $X_r$ denote, respectively, the
$N$, $\Delta(1232)\frac{3}{2}^+$ and the
$N(1440)\frac{1}{2}^+$,
$\Delta(1600)\frac{3}{2}^+$.
With this, $c_{\Delta}=c_{\Delta_r}= 0$, and the coefficients in 
the expansions of Eqs. (\ref{eq:e1},\ref{eq:e11})
are correlated with each other:
\begin{eqnarray}
&&b_{N} \simeq -b_{\Delta} 
\simeq -b_{N_r}\simeq b_{\Delta_r} <~0,
\label{eq:e2}\\
&&c_{N} \simeq -c_{N_r}\simeq b_{N} <~0.
\label{eq:e21}
\end{eqnarray}

Absolute values of coefficients in Eqs. (\ref{eq:e2},\ref{eq:e21})
are $\sim 0.22$ in Refs. \cite{Gershtein,Isgur82}
and $\sim 0.29$ in Ref. \cite{Isgur87}. 
The coefficients $a_X$ are respectively:
$a_N\simeq a_{N_r}\simeq 0.95$, $a_{\Delta}\simeq a_{\Delta_r}\simeq 0.97$ and
$a_N\simeq a_{N_r}\simeq 0.91$, $a_{\Delta}\simeq a_{\Delta_r}\simeq 0.96$.

In this note we present
our results for the 
$\gamma^* N \rightarrow \Delta(1232)\frac{3}{2}^+$,
$N(1440)\frac{1}{2}^+$, and
$\Delta(1600)\frac{3}{2}^+$ transitions obtained within LF RQM
taking into account the configuration mixings from Eqs. 
(\ref{eq:e1}-\ref{eq:e21}).
The results are presented 
in Sections \ref{delta},\ref{roper},\ref{p33} and
summarized in Sec. \ref{summary}.

\section{The $\Delta(1232)\frac{3}{2}^+$ resonance}
\label{delta}

The role of the states $[56',0^+]$ and $[70,0^+]$
in the description
of the $\gamma^* N\rightarrow N$ 
and $\gamma^* N \rightarrow \Delta(1232)\frac{3}{2}^+$
form factors has been considered
and presented
in very detail in Ref. \cite{Aznauryan93}. It  
was shown that with the relations of  Eqs. (\ref{eq:e2},\ref{eq:e21}),
the summary contribution of higher excitation states 
to nucleon form factors 
leads to results that are equivalent
to the results for the pure $[56,0^+]$ state at a different value 
of the parameter $\alpha$
in the radial part of the
wave functions:
\begin{equation}
\Phi\sim exp(-M_0^2/6\alpha^2),
\label{eq:e4}
\end{equation}
where $M_0$ is the invariant mass of three quarks. More specifically,
calculations show that
the results for the nucleon form factors obtained in Ref. \cite{Aznauryan2012}
for the nucleon as pure $[56,0^+]$ state 
with $\alpha=0.37~$GeV
are reproduced for the nucleon from Eq. (\ref{eq:e1})  
with $\alpha=0.33~$ and $0.32~$GeV for the configuration mixings
from Refs. \cite{Gershtein,Isgur82}
and \cite{Isgur87}, respectively.

The states from $[56',0^+]$ do not contribute to the 
$\gamma^* N \rightarrow \Delta(1232)\frac{3}{2}^+$ transition
because $b_{N} \simeq -b_{\Delta}$ (see Eq. (\ref{eq:e2})),
and the state $\frac{1}{2}^+$
from $[70,0^+]$ gives negligible contribution to 
this transition.
Therefore, for $\gamma^* N \rightarrow \Delta(1232)\frac{3}{2}^+$
the difference in the LF RQM predictions,
caused by the admixtures of higher excitation states
in the $N$ and $\Delta(1232)\frac{3}{2}^+$,
is determined only by the replacement
$\alpha=0.37~\rightarrow~0.32$ (or $0.33$) GeV.

The predictions for the $\gamma^* N \rightarrow N$ form factors and 
the $\gamma^* N \rightarrow \Delta(1232)\frac{3}{2}^+$ and
$\gamma^* N \rightarrow N(1440)\frac{1}{2}^+$ transitions 
were made in Refs.
\cite{Aznauryan2012,Aznauryan2015} under the assumption
that in addition to the three-quark (3q) contribution, these transitions 
contain contributions, which are produced by meson-baryon
interaction. 
The nucleon electromagnetic form factors were described
by combining $3q$ and pion-nucleon loops contributions.
With the pion-nucleon loops evaluated according to
the LF approach of Ref. \cite{Miller}, the following form of the nucleon
wave function has been found:
\begin{equation}
|N>=0.95|3q>+0.313|\pi N>.
\label{eq:eq1}
\end{equation}

For the 
$\Delta(1232)\frac{3}{2}^+$ and
$N(1440)\frac{1}{2}^+$,  
the weights of the $|3q>$ component in the expansion
\begin{equation}
|X>=c_{3q}(X)|3q>+...~~~
\label{eq:eq3}
\end{equation}
were found 
from experimental data on
the $\gamma^* N \rightarrow \Delta(1232)\frac{3}{2}^+$ and
$\gamma^* N \rightarrow N(1440)\frac{1}{2}^+$  
transitions
assuming that at $Q^2 > 4~$GeV$^2$
these transitions
are determined only by the first term in Eq. (\ref{eq:eq3}).

\begin{figure}[htp]
\begin{center}
\includegraphics[width=7.6cm]{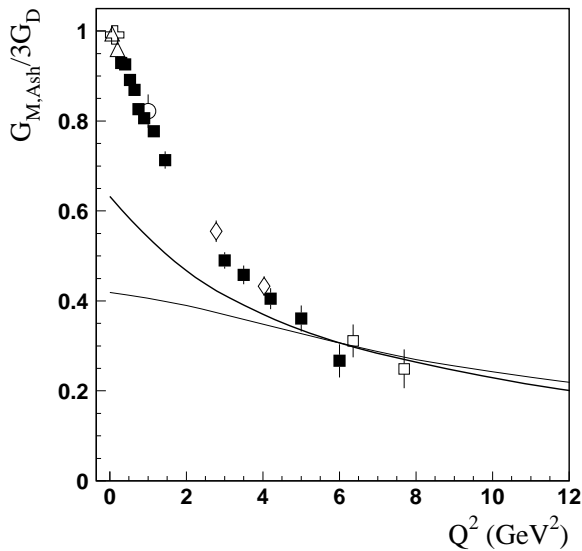}
\caption{\small
The form factor $G_{M,Ash}(Q^2)$
for the  $\gamma^* p \rightarrow~\Delta(1232)\frac{3}{2}^+$
transition relative to $3G_D$: $G_D(Q^2)=1/(1+\frac{Q^2}{0.71GeV^2})$.
The full boxes are the CLAS data extracted
in the analysis of Ref. \cite{Aznauryan2009}.
The results from other experiments are: 
open boxes \cite{Vilano},
open triangles \cite{Stave2006,Sparveris2007,Stave2008},
open cross \cite{Mertz,Kunz,Sparveris2005}, open rhombuses \cite{Frolov},
and open circle \cite {KELLY1,KELLY2}.
The thin solid curve presents the LF RQM predictions from Ref. \cite{Aznauryan2015}
obtained for
the $N$ and $\Delta(1232)\frac{3}{2}^+$ taken as pure states 
in the multiplet $[56,0^+]$.
The thick solid curve corresponds to the LF RQM results 
when the admixtures of 
higher excitation states in the $N$ and $\Delta(1232)\frac{3}{2}^+$ 
are taken into account 
with the weights from Refs. \cite{Gershtein,Isgur82,Isgur87}.  
\label{gm_ash}}
\end{center}
\end{figure}

The predictions for the magnetic-dipole 
$\gamma^* p \rightarrow~\Delta(1232)\frac{3}{2}^+$
form factor  in the Ash convention \cite{Ash}
are presented in Fig. \ref{gm_ash}. 
In the case, when
the $N$ and $\Delta(1232)\frac{3}{2}^+$ are pure $[56,0^+]$ states,
the predictions
are given by the thin solid curve; they 
coincide with 
the results of Ref. \cite{Aznauryan2015}. 

The thick solid curve corresponds to the results when the admixtures of 
higher excitation states in the $N$ and $\Delta(1232)\frac{3}{2}^+$ are taken into account.
These admixtures 
increase the $3q$ contribution to the 
$\gamma^* N \rightarrow \Delta(1232)\frac{3}{2}^+$
magnetic-dipole
form factor at $Q^2=0$ from $42\%$ to $63\%$.
The weight of the $|3q>$ component
in the $\Delta(1232)\frac{3}{2}^+$  
is, respectively,
$c_{3q}(\Delta)=0.84\pm 0.04$
and
$c_{3q}(\Delta)=0.91\pm 0.04$
for the admixtures of 
higher excitation states 
from Refs. \cite{Gershtein,Isgur82}  
and \cite{Isgur87}. 

We note, that 
both predictions for the $3q$ contribution to the 
$\gamma^* N \rightarrow \Delta(1232)\frac{3}{2}^+$
magnetic-dipole form factor are within limits 
obtained in the dynamical reaction model
\cite{Lee,Lee1}, where the bare contribution, that can be associated with
the $3q$ contribution, gives at $Q^2=0$ 
about $40-70\%$ of the total magnetic-dipole
form factor.

For the ratios $R_{EM}$ and $R_{SM}$, 
the admixtures of 
higher excitation states in the $N$ and $\Delta(1232)\frac{3}{2}^+$,
do not affect the results obtained 
in Ref. \cite{Aznauryan2015}.

\section{The $N(1440)\frac{1}{2}^+$ resonance}
\label{roper}
Similar to the electroexcitation of the $\Delta(1232)\frac{3}{2}^+$, the second terms 
in the expansions of Eqs. (\ref{eq:e1},\ref{eq:e11})
do not contribute to the $\gamma^* N \rightarrow N(1440)\frac{1}{2}^+$ transition
because $b_{N} \simeq -b_{N_r}$. In addition, the states 
from $[70,0^+]$ give negligible contribution to this transition.
Therefore, for $\gamma^* N \rightarrow N(1440)\frac{1}{2}^+$ too,
the difference in the LF RQM predictions,
caused by the configuration mixings, is determined only by the replacement
$\alpha=0.37~\rightarrow~0.32$ (or $0.33$) GeV.

\begin{figure*}[htp]
\begin{center}
\includegraphics[width=12.0cm]{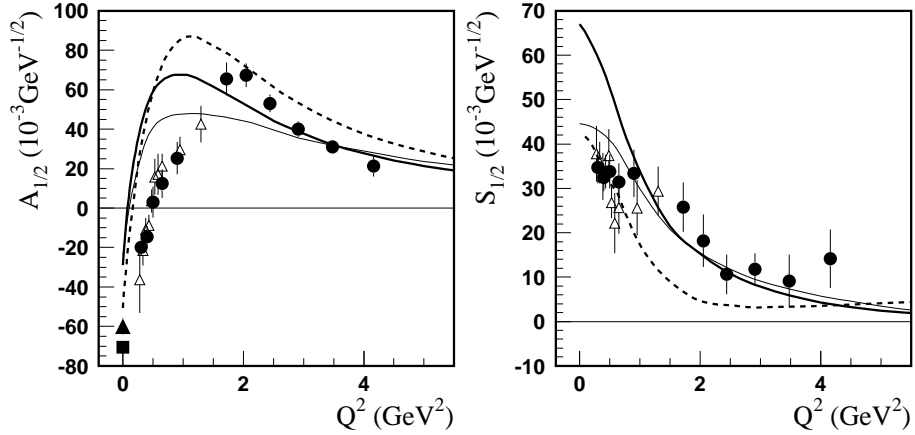}
\vspace{-0.1cm}
\caption{\small
The $\gamma^*p\rightarrow N(1440)\frac{1}{2}^+$
transition helicity amplitudes.
Solid circles and open triangles are the amplitudes extracted 
from the CLAS data on $\pi N$  \cite{Aznauryan2009} 
and $\pi^+\pi^- p$ \cite{Mokeev,Mokeev1} electroproduction 
off the proton. 
The full box at $Q^2=0$ is
the amplitude extracted from the CLAS $\pi$ photoproduction
data \cite{Dugger}.
The full triangle at $Q^2=0$ is
the Review of Particle Physics (RPP) estimate \cite{RPP}.
The thin solid curves present the LF RQM predictions from Ref. \cite{Aznauryan2012} 
obtained for
the $N$ and $N(1440)\frac{1}{2}^+$ taken as pure states in
the multiplets $[56,0^+]$ and $[56',0^+]$.
The thick solid curves correspond to the LF RQM results when the configuration
mixings in the $N$ and $N(1440)\frac{1}{2}^+$ are taken into account
with the weights from Refs. \cite{Gershtein,Isgur82,Isgur87}.
The dashed curves are the results obtained within 
Dyson-Schwinger equations (DSE) in QCD \cite{Roberts,Roberts1}.
\label{p11}}
\end{center}
\end{figure*}

The predictions for the 
$\gamma^* N \rightarrow N(1440)\frac{1}{2}^+$ 
helicity amplitudes
are presented in Fig. \ref{p11}.
In the case, when
the $N$ and $N(1440)\frac{1}{2}^+$ are pure states in $[56,0^+]$ 
and $[56',0^+]$, the predictions are given by the thin solid curves; they
coincide with the results of Ref. \cite{Aznauryan2012}.

The thick solid curves correspond to the results when configuration
mixings in the $N$ and $N(1440)\frac{1}{2}^+$ are taken into account.
These mixings clearly improve the agreement with 
experiment for the transverse $A_{1/2}$ amplitude above $1.5~$GeV$^2$, 
while the behavior of this amplitude below $0.6~$GeV$^2$ remains unchanged.
The weight of the $|3q>$ component
in the $N(1440)\frac{1}{2}^+$ 
is, respectively,
$c_{3q}(N_r)=0.91\pm 0.05$
and $c_{3q}(N_r)=0.95\pm 0.05$.
with the configuration mixings from Refs. \cite{Gershtein,Isgur82}
and \cite{Isgur87}. 

In Fig. \ref{p11}, we also present the predictions obtained
within DSE's in QCD \cite{Roberts,Roberts1},
which allow most direct connection between quark-quark interaction
of QCD and hadron observables. We note remarkable agreement
between $Q^2$ dependences
of the transverse amplitude $A_{1/2}$ 
obtained within DSE's
and in the LF RQM 
with configuration mixings
taken into account. 

\section{The $\Delta(1600)\frac{3}{2}^+$ resonance}
\label{p33}

In contrast with the $\Delta(1232)\frac{3}{2}^+$ and $N(1440)\frac{1}{2}^+$, 
configuration mixings have a very strong impact on the 
results for the $\Delta(1600)\frac{3}{2}^+$.
The reasons are following:

(1) The contribution to $\gamma^* N\rightarrow \Delta(1600)\frac{3}{2}^+$ 
corresponding to the unmixed $|N,3q>=[56,0^+]$ and $|\Delta_r,3q>=[56',0^+]$ is 
suppressed compared to the contributions from the
transitions between
$|N,3q>=[56,0^+]$ and $|\Delta_r,3q>=[56,0^+]$, and
$|N,3q>=[56',0^+]$ and $|\Delta_r,3q>=[56',0^+]$.
For example,  for the helicity amplitudes in the units
of $10^{-3}$GeV$^{-1/2}$ at $Q^2=0$ we have:
\begin{eqnarray}
A_{1/2}= -49.7(1+3.9a_{_{\Delta_r}}b_{_N}+4.3a_{_N} b_{_{\Delta_r}}),
\label{eq:e41}\\
A_{3/2}= -54.6(1+5.9a_{_{\Delta_r}}b_{_N}+6.4a_{_N} b_{_{\Delta_r}}),
\label{eq:e42}\\
S_{1/2}= -16.8(1+4.8a_{_{\Delta_r}}b_{_N}+5.5a_{_N} b_{_{\Delta_r}}).
\label{eq:e43}
\end{eqnarray}

In addition, in contrast with the
$N(1440)\frac{1}{2}^+$, for the $\Delta(1600)\frac{3}{2}^+$ we have 
$b_{N}\simeq b_{\Delta_r}$ (see Eq. (\ref{eq:e2})).

From Eqs. (\ref{eq:e41}-\ref{eq:e43}), it follows that at $Q^2=0$ the
$\gamma^* N\rightarrow \Delta(1600)\frac{3}{2}^+$ 
helicity amplitudes change their sign.
With increasing $Q^2$, the relative values of the
contributions caused by the configuration mixings  
become smaller, and the 
$\gamma^* N\rightarrow \Delta(1600)\frac{3}{2}^+$ 
amplitudes don't change sign.

\begin{figure*}
\begin{center}
\includegraphics[width=16.8cm]{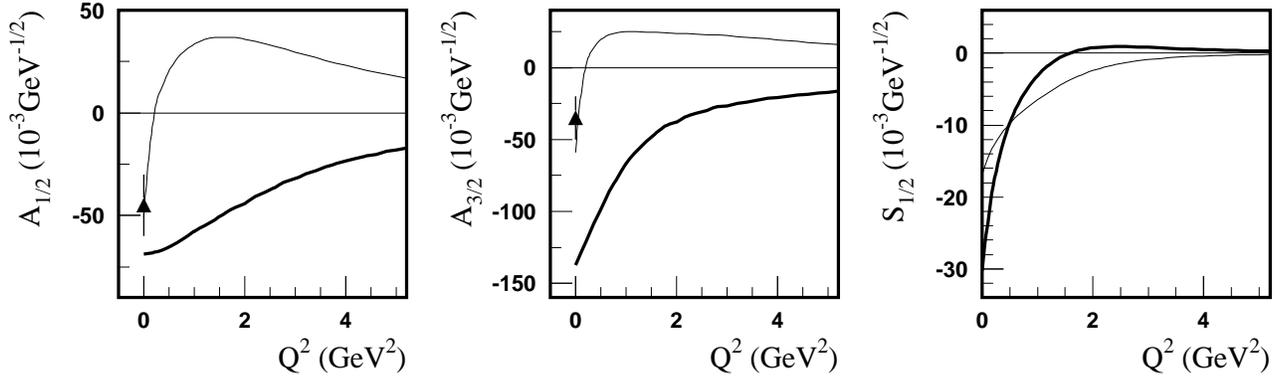}
\vspace{-0.1cm}
\caption{\small
Helicity amplitudes
for the $\gamma^* p \rightarrow~\Delta(1600)\frac{3}{2}^+$
transition.
The full triangles at $Q^2=0$ are the RPP estimates \cite{RPP}.
The thin solid curves present the LF RQM predictions when
the $N$ and $\Delta(1600)\frac{3}{2}^+$ are pure states from the multiplets $[56,0^+]$ 
and $[56',0^+]$.
The thick solid curves correspond to the results when the configuration
mixings in the $N$ and $\Delta(1600)\frac{3}{2}^+$ are taken into account
with the weights from Refs. \cite{Gershtein,Isgur82,Isgur87}.
\label{p33_1600}}
\end{center}
\end{figure*}

(2) For the reasons given in point (1), the additional 
sign in the helicity amplitudes related to the
$\pi N \Delta(1600)\frac{3}{2}^+$ vertex is changing. Let us remind,
that this sign has been found in Refs. \cite{Aznauryan2015,Aznauryan2007}
in the approach based on partial conservation of axial current (PCAC)
and is determined by the expression:
\begin{equation}
I_{NA}\equiv \int\frac
{(m_q+M_0x_a)^2-\mathbf{q}_{a\perp}^2}
{(m_q+M_0x_a)^2+\mathbf{q}_{a\perp}^2}
\Phi_N(M_0^2)
\Phi_A(M_0^2)d\Gamma,
\label{eq:e44}
\end{equation}  
where $A$ denotes the states $N$, $\Delta(1232)\frac{3}{2}^+$, 
$N(1440)\frac{1}{2}^+$,
and $\Delta(1600)\frac{3}{2}^+$. In Eq. (\ref{eq:e44}), 
$m_q$ is the constituent quark mass, $x_a$ is the fraction of the
initial nucleon momentum
carried by the active quark, $\mathbf{q}_{a\perp}$ is the transverse
momentum of this quark, and $d\Gamma$ is the phase space volume of the quarks.
These quantities are defined in more detail in 
Refs. \cite{Aznauryan2015,Aznauryan2007}.
For the reasons mentioned in point (1),
the expression in Eq. (\ref{eq:e44}) changes its  sign
for the $\Delta(1600)\frac{3}{2}^+$, while 
for the $N$, $\Delta(1232)\frac{3}{2}^+$, 
and $N(1440)\frac{1}{2}^+$ its sign remains unchanged.
Therefore, finally at $Q^2=0$ the
$\gamma^* N\rightarrow \Delta(1600)\frac{3}{2}^+$ 
helicity amplitudes change their sign twice and remain negative.
At large $Q^2$, these amplitudes 
change their sign once.
The $Q^2$ evolution of the predictions for the helicity
amplitudes is given in Fig. \ref{p33_1600}.

\section{Summary}
\label{summary}

We have shown that the configurations mixings
which follow from QCD-inspired interquark forces
with the weights found in Refs. \cite{Gershtein,Isgur82,Isgur87}
have significant impact on the LF RQM predictions for
the electroexcitation of the
$\Delta(1232)\frac{3}{2}^+$ and
$N(1440)\frac{1}{2}^+$, and result in qualitative changes of the 
helicity amplitudes predicted for the 
$\gamma^* N\rightarrow \Delta(1600)\frac{3}{2}^+$ transition.
 
For the $\Delta(1232)\frac{3}{2}^+$, 
the admixtures of
higher excitation states in the $N$ and 
$\Delta(1232)\frac{3}{2}^+$ 
increase the $3q$ contribution to the
$\gamma^* N \rightarrow \Delta(1232)\frac{3}{2}^+$
magnetic-dipole form factor, in particular 
at $Q^2=0$, the $3q$ contribution to this form factor grows from $42\%$ to $63\%$.
The predictions for the ratios $R_{EM}$ and $R_{SM}$
remain unchanged.

For the $N(1440)\frac{1}{2}^+$, incorporating 
the configuration mixings in the $N$ and
$N(1440)\frac{1}{2}^+$  leads to better agreement with experiment 
for the transverse helicity amplitude above $1.5~$GeV$^2$.
The predictions 
for this amplitude below $0.6~$GeV$^2$,
including zero-crossing at $Q^2\simeq 0.1~$GeV$^2$,
remain unchanged. 

For the $\Delta(1600)\frac{3}{2}^+$, 
the specific behavior of the transverse helicity amplitudes
with zero-crossing near $Q^2=0.2~$GeV$^2$, obtained earlier
for the pure $N$ and $\Delta(1600)\frac{3}{2}^+$ 
from the multiplets $[56,0^+]$
and $[56',0^+]$, disappears
when configuration mixings in these states are taken into account.
The configuration mixings result also in the change of the additional sign
related to 
the $\pi N \Delta(1600)\frac{3}{2}^+$ vertex. As a result,
both transverse amplitudes become negative at all $Q^2$
with absolute values that slowly decrease with increasing $Q^2$. 

{\bf Acknowledgments}.
We are grateful to C. Roberts and V. Mokeev for useful
correspondence.
This work was supported by
the U.S. Department of Energy, Office of Science,
Office of Nuclear Physics, under Contract
No. DE-AC05-06OR23177, and the National Science Foundation, State
Committee of Science of the Republic of Armenia, Grant No. 15T-1C223.

\end{document}